\documentclass[11pt]{article}
\pdfoutput=1  % arXiv: force pdfLaTeX engine

\usepackage[T1]{fontenc}
\usepackage[utf8]{inputenc}
\usepackage[margin=1in]{geometry}
\usepackage{graphicx}
\usepackage{array}
\usepackage{caption}
\usepackage[expansion=false]{microtype}
\usepackage[hidelinks]{hyperref}
\usepackage{xurl}
\usepackage{orcidlink}

\newenvironment{references}{%
  \section*{Bibliography}%
  \mbox{}\par\vspace{-\baselineskip}% absorb \@afterheading so entry 1 also hangs
  \small
  \setlength{\parindent}{-2em}%
  \setlength{\leftskip}{2em}%
  \setlength{\parskip}{0.5em}%
  \par
}{\par}

\title{\textbf{The Market in the Model:}\\[0.2em]
\textbf{Latent Diffusion as Neural Economy}}

\author{%
  Eryk Salvaggio\,\orcidlink{0000-0002-9103-3734}\\[2pt]
  \small Cambridge Digital Humanities, University of Cambridge\\
  \small Machine Visual Culture Research Group, Max Planck Institute\\[2pt]
  \small \texttt{es2082@cam.ac.uk}
}

\date{\small Preprint, June 17, 2026}

\begin{document}

\maketitle

\begin{abstract}
\noindent Valuable critique of generative image models within visual culture and the humanities has emphasized the role of datasets in shaping the images they produce. Yet, close studies of the ideological positions embedded into the mechanism of the models have been neglected, leaving them imagined as ``black boxes.'' In a bid to expand, rather than replace, dataset critique, this paper examines the mechanisms of the latent diffusion model in terms of the problems they were brought in to solve on behalf of computer vision engineers, and the decisions each component was tasked with automating. I interpret that ensemble through the histories of its parts and the theory of vision the system inscribes into every generated image. Drawing on Impett and Offert's notion of \emph{neural exchange value}, I offer this analysis to argue that the model operates as a \emph{neural economy}: a contained symbolic system that abstracts social communication into commensurable vectors as it transfers the social sphere into parcels for sale. Tracing the training and generation pipelines component by component reveals what each operation displaces, and how it further entrenches the logics of platform and attention economies over social communication. The paper warns that any critique fixated exclusively on copyright and commodity defenses risks reaffirming the very fetishism the model produces, and argues instead for centering social exchange.

\end{abstract}

\section*{Introduction}

The architecture of a latent diffusion model is not a single object but a discursive process: a series of automated decisions that inscribe themselves into an image. Each image is the product of an ensemble of scrapers, classifiers, rankers, filters, and feedback systems, often built for some other purpose prior to their role in image production. I read this ensemble through the history of these components, which engineers fused into the original latent diffusion model, and examine the accidental theory of vision and images this assemblage writes into the generated image.

In a 2021 paper, engineers from the Computer Vision \& Learning Group at LMU Munich presented the Latent Diffusion model (LDM) architecture that would soon be integrated into Stable Diffusion. The most influential open-source image-generation model for consumer access after OpenAI's DALL-E 2, Latent Diffusion architecture continues to serve as a foundational reference for image models, a template that other systems have expanded or modified while preserving most of its core structural assumptions and logics across Stable Diffusion, Midjourney, and other generative AI image systems. It also made its weights and documented components available, offering insight into the system and the logics that shaped it (Rombach 2021).

Latent Diffusion was developed with support from the startup RunwayML. Financial incentives shaped the design, guiding which datasets and components to include: the resulting systems were a series of processes for recognizing images, scraping web pages, and collecting aesthetic feedback under the banner of efficiency. Automating these decisions reflected an ideological position on what the mechanism should emphasize, value, or discard. Beyond the dataset, these decisions shape what is passed from one component to the next, positions which compound within the model weights during training. The generated image becomes a surface on which these values are inscribed, presented to the user as if it were a spontaneously produced result of some internal creative deliberation.

But the reality is quite different. I situate the AI training regime as a system designed to transfer language and images, vessels of communicating \emph{social} exchange, into a new economy and definition of value. That transfer occurs first through total abstraction of the training images, subjected to mechanisms of commensuration and exchange. This is what Impett and Offert define as \emph{neural exchange value}, a non-market form of abstraction in a model that establishes equivalencies between symbols and vectors within an AI system (2026). For Impett and Offert, every model takes the form of a market: its own attempt to universalize a training set into a complete map of the world. Each model builds a map through abstraction, and no two of these maps are compatible --- each treats its own version as the whole. I examine the LDM as another attempt at creating forms of a world through its own form of abstraction.

Pasquinelli writes that ``the inner code of AI is constituted not by the imitation of biological intelligence but by the intelligence of labor and social relations'' (2023, 13). Building on this, I read the latent diffusion model as running on a logic of resources and commodities, siphoned from the congealed labor of social exchange --- a logic that produces an economy operating as a market. But this alien economy is not anchored to the exchange of commodities, in that it does not sell \emph{things}, even virtual ones: it sells access to their production. Nor is it an economy of social exchange, as it is not designed to \emph{communicate} to the user when it constructs an image. It is an intermediary for selling a model of the world which becomes the basis of a service. I call this intermediate stage a \emph{neural economy}: a contained symbolic system in which neural exchange value circulates. It operates on data commodified for no direct social or economic use, existing solely to construct a model of representation to produce media, code, text, and so forth. In sum, the model is a conduit, a pipeline for abstraction facilitating the \emph{transfer} of social communication \emph{into} a commodity form.

\section*{Theoretical Framework and Methodology}

Image models are classification systems running in reverse, as the compression used to name and describe images becomes an outline for generating new ones. These models decompress billions of images and return them as a single legible output, and to do so they must refabricate any specificity that compression discards. Fazi describes generative AI as ``a world within''--- a representational reality with its own internal consistency, not accountable to the outside world (2024, 47). This paper examines the historical and financial conditions that structure and govern that world within the latent diffusion model, and how they came to be. I use \emph{model architecture} to name what Latour's black box (1999, 304) typically obscures and what Star's infrastructure (1999) describes: the accumulated logics, assumptions, and historical residues that shape system behavior. A model's components are diverse in origin, but structurally fused (Shoemaker 2026). When the seams disappear, the cluster is read as a unified object: \emph{the model}. I resist reducing the model to a single ``neural network,'' and instead examine the multiple systems it contains, each designed to meet specific goals: even the most basic latent diffusion model is not \emph{a} neural network but at least three. This analysis isolates them by the task each network is asked to do, rather than by tracing internal weights and embeddings.

To grasp the underlying mechanisms of these systems, I define three types of exchange:

\textbf{Social exchange value} is the value an image or text carries through social use: its work as a shared reference others can interpret. The connection between image and world is social --- the image is a sign that communicates. A photograph of a pet dog posted to an online forum circulates for this kind of exchange. The value is the social exchange.

\textbf{Neural exchange value} is oriented toward \emph{commensuration}: the reduction of specific, situated things into exchangeable units whose particularity is discarded through compression ``from the symbolic to the geometric'' (Impett and Offert 2026, 15) as it moves between a specific image and its vector-space representation. Consider the dog photo in the vector space: no longer a specific dog nor even an image per se, but weights and vectors that influence the position a dog image holds relative to other images. Social relations appear as relations between vectors that connect images to text.

\textbf{Commodity exchange value} is accumulated through abstraction to a market and congealed into the form of a commodity. For example, how much is a photograph of your dog worth in financial terms, or the valuation of the model trained on it? How much is a synthetic image of a dog worth? This extends onward into the political and financial economy of commodities. Social relations between people appear as relations between things.

Commodity logics have shaped the policy conversation and analytical lens on training data, particularly through copyright. But this framing runs into trouble when neural and financial value are treated as a single mechanism. A single image is a minuscule fraction of a dataset, rendered into math inside a model; and digital outputs lack the traditional sources of commodity value, such as scarcity.

Neural exchange value is not a commodity in the traditional sense. But it behaves like one, because it shares an assumption from markets: that everything is comparable to everything else, if you can find the right number. Impett and Offert find this at work in the vector space, and trace it to Geoffrey Hinton, who himself drew on Marx's commodity exchange to theorize the computational mechanism that gave rise to generative AI models. To enter a model, a thing must first be \emph{adaptable to universal commensurability}. This is the ground on which generative AI's machine aesthetics and politics were seeded: everything in the vector space is designed to be quickly transformed into something else.

The model inherits an ideology: what is useful to it is only what can be converted, and converting what was once unconvertible is the work of the industry that profits from these models. Sohn-Rethel (1978) argues that this habit --- lining unlike things up on a single scale --- is not natural but learned from the market. The mechanisms that turn words and images into vectors carry it forward, treating the world as though everything in it were already exchangeable. Whatever falls outside this system is discarded or flattened. The model alone does not accumulate capital, but renders an ever larger, ever more commensurable picture of the world as a means toward capital. All forms of life in this ideology are condensed into data, and in turn, all data congealed into models. Whenever a company proposes a new way to turn some untamed part of the world into data, this is neural exchange value; whenever that data is operationalized, it is translated into commodity value.

As such, what can be made commensurable, and what can be discarded, are the animating features automated beneath the sealed system of the AI model. We can read it in images the model discards during training, what the autoencoder is trained to recognize as photorealistic, the mechanisms that set the thresholds of value for the model's desired users. It is an ideology encoded into a sequence of operations, and that sequence organizes the generated image through that ideology.

The commercial market is the first mechanism for that conversion. Training data is abstracted from online conversation and image-sharing, already shaped by platform capitalism (Crawford and Paglen 2021), into a series of numerical weights aimed at reproducing the patterns of that speech and image-sharing. That masks an ideological orientation: no single piece of data in the model has any inherent financial value but is valuable to the model only for what it allows to be \emph{reduced and located}. Popular resistance to AI often leans into financial instruments such as copyright law to protect artists from this transfer. Yet so much of what circulates online also holds \emph{social exchange value}, anchored in community participation and conversation, rather than pure financial value. The model's output then re-circulates as a product of the ever-tightening compression of communication and commodification, while flattening the difference between the world within the model and the world of media and communication from which it siphons.

Software has always encoded and re-executed political assumptions, often without any single actor intending it (Chun 2013). We can follow \v{Z}i\v{z}ek to ask not why actors behave as they do, but \emph{why the system took this specific form} (1989, 11). I will look beyond the symptom of the generated image to trace the logic of available components that determined the form of the latent diffusion model, and by extension, the images they produce. Reading the generated image as a cultural artifact means accounting for this absence of the world to see what has replaced it (Salvaggio 2022; Steyerl 2023). To understand the surface, we turn to the substrate.

\section*{A Two-Phase Architecture}

The core innovation behind the Latent Diffusion Model was to run the diffusion process --- introduced by Sohl-Dickstein et al. (2015) --- in a compressed latent space rather than in pixels (Rombach 2021). A pre-trained autoencoder first compresses images into a lower-dimensional space; the diffusion process then runs there, not at full resolution. This cut the computational cost enough to run on consumer hardware, the commercially viable system the team was after.

Several distinct components are embedded within this model, and each operation adds a layer of ideological residue: a scraper gathers training data (Common Crawl); a system pairs images with text (CLIP); a compressor reduces images to representations (the autoencoder); a denoiser recovers images from noise (the U-Net); a mechanism weighs the image against prompt (CFG). What follows is a deeper analysis of these components. The training section traces how this logic enters the model. Following that, the generation section traces it as images are made.

\section*{Training}

\begin{quote}
\itshape
The ideology behind the diffusion model facilitates a wholesale transfer of the social sphere into commodity through neural exchange value.
\end{quote}

\subsection*{CLIP}

Contrastive Language-Image Pretraining (CLIP) is a system that learns to match images with the text that describes them. Trained on hundreds of millions of image-caption pairs scraped from the web, it produces a numerical score for how well any image matches any caption --- a score the rest of the pipeline uses to filter, sort, and evaluate. CLIP was trained on the World Wide Web, where social exchange occurs without financial compensation, because social participation in online communities is its own goal. Terranova (2000) describes the internet as a site for collectivizing ``immaterial labor,'' what Lazzarato defines as ``the work of defining cultural standards, tastes, and public opinion'' (1996, 133), producing text whose value lies in communication rather than in individual economic benefit.

Nonetheless, a market arose around online exchange: the platform economy, which sells digital space based on attention. The price of advertising on a page reflects how many people view it. A former Google executive founded Common Crawl (CC) to help startups compete with Google's search index by archiving and measuring the Web (Reisner 2025). CC ranks pages using `Harmonic Centrality,' a weighted tally of inbound links (Baack 2024). Advertising-supported content, e-commerce, social media, and stock image libraries dominate the data. It is heavily biased toward English-language, Western sources and holds substantial harmful content (Birhane 2021; Luccioni and Viviano 2021). Applied to the scraped social exchanges of text and images, CC's Harmonic Centrality reflects the political economy of Cognitive Capitalism found in Google PageRank: ``it is not simply an apparatus of surveillance or control, but a machine to capture living time and living labour and to transform the common intellect into \emph{network value}'' (Pasquinelli 2009, 2).

The CC dataset was created to answer a question: \emph{what do people link to online?} Repurposed as the training data for the Latent Diffusion model, the question posed to CC shifted to: \emph{what does the representable world look like?} In the gap between the two, ideology slipped in. The representable world looked like whatever had the most inbound links, described by whatever caption the commercial web placed alongside it. The Latent Diffusion team pushed these biases further by relying on a derivative dataset of text and images, LAION-400M, which filtered CC using CLIP (Birhane 2023). CLIP was trained on 400 million image-text pairs (Agarwal 2021). It evaluates images based on statistical proximity to similar captions rather than any unique or distinguishing context: the goal is generality, rather than specificity, and variety is used mostly to push the boundaries of the category assigned to it. It also solved a labor problem: labeling datasets at scale requires enormous human effort. ImageNet (Deng et al. 2009) required 49,000 workers to annotate 14 million images (Li and Deng 2017; Denton et al. 2021). CLIP replaced that judgment without hiring new workers --- instead embedding the labor of earlier human annotators into its weights (Radford 2021).

One ideological orientation of CLIP is visible in what it filters out: images without captions, or with captions that deviate from English-language text conventions. This further entrenched Common Crawl's initial biases, surfacing only image-text pairs that CLIP could reliably identify. Browne describes such patterns of algorithmic omission as ``a logic of prototypical whiteness'' (2015) reflecting the racial makeup of Silicon Valley managers, perpetuating unexamined gaps which are found throughout the world (Qadri 2023; Ghosh 2024; Alenichev 2024; Turk 2023). Birhane (2021) identifies the main driver of this impulse in LAION as an example of ``scale thinking'': the assumption that efficiency can be endlessly expanded without changing inputs or infrastructure (Hanna and Park 2020, 1). CLIP was an attempt to scale without human labor, compressing the world into a narrow and filtered set of symbolic representations that reflected these ideological priorities.

CLIP makes visual and textual data interoperable by encoding them into a shared vector space. The vector space is a geometry: every image and every caption becomes a point in a high-dimensional grid, and proximity in the grid stands in for similarity in meaning. Two images of dogs sit near each other; a dog and a bicycle sit far apart. The model never sees images or words as such --- only their coordinates. CLIP moves these image and text labels into this vector space to create a computationally interpretable measure of similarity (Impett and Offert 2026, 21). In doing so, it completes a transformation Meyer finds in the broader visual economy: style ``ceases to be a historical category and becomes a pattern of visual information to be extracted and monetized'' (2023, 107). CLIP introduces an enforcement mechanism: it doesn't merely treat style as a pattern of difference between individual images, but measures and encodes the distinction at the level of the vector. Instead of setting a price, it sets a position.

What Common Crawl commodified by measuring links to an item on the Web, CLIP fixes through visual-textual proximity: the assignment of social and economic value is driven by the logic of the attention economy. The shift from \emph{social exchange value} (the images people write and say online) to \emph{neural exchange value} (the use of those images to set up a topographic map of concepts in the latent space) reinforces and commodifies the structural logics found across digital communication platforms. It does so to create a system of repeatable references for the improvement of the model's world-inside.

\subsection*{The Autoencoder}

The next stage is the autoencoder, which is meant to be the model's approximation of sight: a neural network that compresses images into a compact numerical representation and then reconstructs them. The encoder half does the compressing; the decoder half does the reconstructing. Whatever is extracted from this compression is what the system can later reproduce. Something is always lost in compression, so the decoder can only reconstruct a plausible-seeming substitute.

How does the machine determine if an image is plausible? Two mechanisms are borrowed from prior systems to assess that resemblance, each with its own criteria for assessment. First, \emph{perceptual loss} measures how closely the reconstruction resembles the original. Perceptual measurement was built on the ImageNet dataset to preserve the quality of streaming video and image compression, but draws from histories of compression dating back to the telephone, where it was a strategy to maximize what can occur within the limits of existing architecture for the sake of expanding profit (Sterne 2012, 33). Second, a patch-based adversarial objective from pix2pix asks, for every $70\times70$-pixel window, ``Does this look real?'' (Isola 2016).

The perception replaced here is a subjective act shaped by the perceiving body, its cultural context, and a history of vision. Perceptual loss mechanisms replace that anchor of bodily perception by substituting statistical proxies for missing details (Zhang et al. 2018, 586). The PatchGAN discriminator makes the same substitution at the local level: a classifier's mathematical thresholds define local realism built to stand in and anticipate the human eye (Demir 2018). The autoencoder thus replaces embodied perceptual judgment --- the capacity to interpret what an image represents --- with a model built by comparing and contrasting image and text data from the CLIP-filtered corpus.

Perceptual loss asks whether the reconstruction resembles the original; the PatchGAN discriminator asks whether local details look real. But the original data are CLIP-filtered images, and their assessment is calibrated against images that the commercial web has already dictated are worthy of measurement. This sorting logic doesn't merely persist in the autoencoder because of the training data; the autoencoder concentrates it into the blurry starting point from which all latent diffusion models produce their final image.

The autoencoder is supposed to stand in for a perceptual subject position, but its capacity for evaluation and comparison is entirely shaped by the commercial web's visual hierarchy. This is a sneaky ideological operation: the autoencoder occupies the structural position of a perceiving eye without meeting that position's subjective demands. The discernment of a human body is replaced by the quantified crudeness based on what passes the commercial web's sorting logic. The crude decision is inscribed in vector geometry, equating perception with the act of evaluating plausibility -- but at best, it asks: \emph{does this image look like an image}? At each step, the picture of a dog becomes more dog-like because it is being compared against an absence of dog, represented by blurry or noisy patches remaining in the image.

At each step of generation, the image is defined against an absence of representational structure, what we may call digital noise. Every generated image begins with a random seed, the source of computational ``randomness'' within the model that produces variety in the image output. Prompting the model with the exact phrase `Gaussian noise' does not reveal the seed, because the model must complete its sequence. Instead, the prompt forces the model to walk from noise toward a specific visual reference that it cannot produce, because it has been filtered out.

Gaussian noise distributions resemble millions of arbitrarily colored pixels in a jpeg. Noise in the Latent Diffusion Model had no structure for perceptual loss to measure, no local texture resembling anything in the training data, and did not pass CLIP to enter the training data because CLIP is structurally looking for its opposite. The model has only one direction: away from the structureless and toward a plausible image. Asked to represent the impossible, it generates the substrate of visual plausibility, but, to the human eye, it depicts nothing at all. These images pass the architecture's perceptual criteria, but not ours. They produce the surface quality of images without the content. The resulting image is, in one sense, an architectural failure: it cannot represent what is unrepresented. Yet, it is \emph{also} an automated conversion of computational noise into a structured, recognizable commodity: the image.

\begin{center}
\begin{tabular}{cc}
\includegraphics[width=0.30\textwidth]{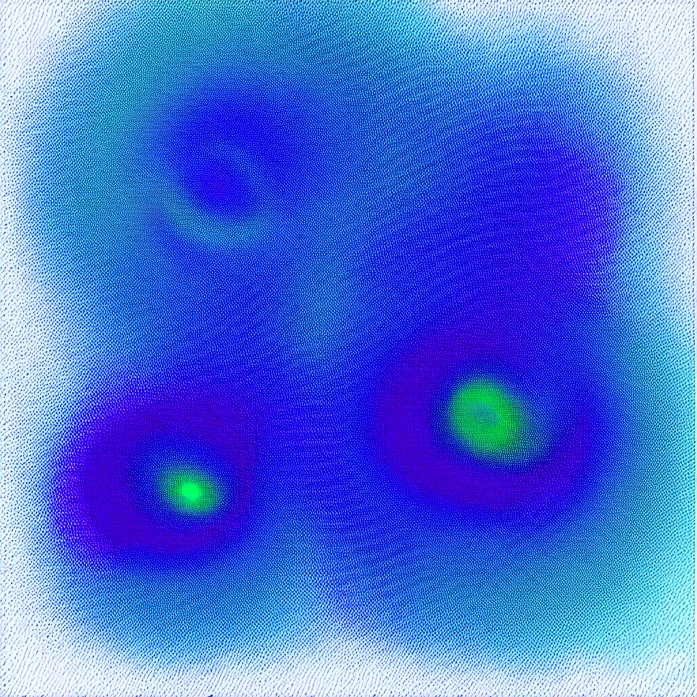} &
\includegraphics[width=0.30\textwidth]{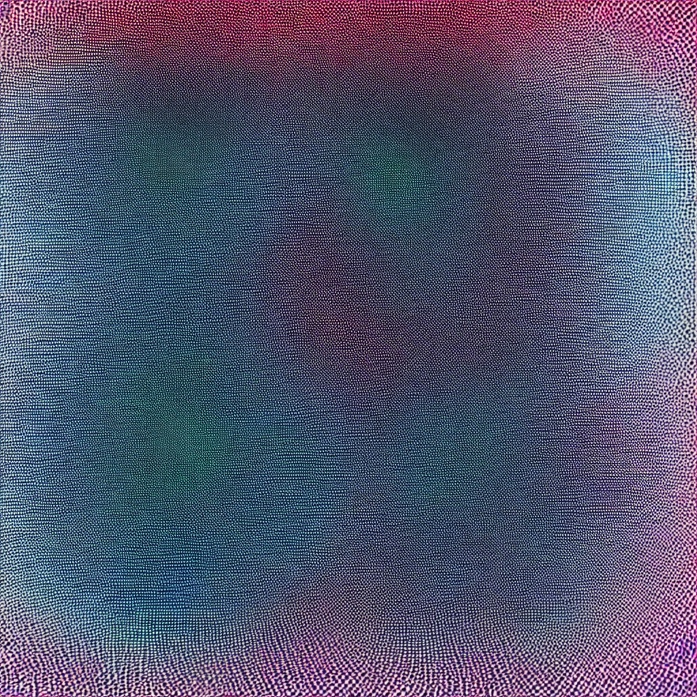} \\[2pt]
\includegraphics[width=0.30\textwidth]{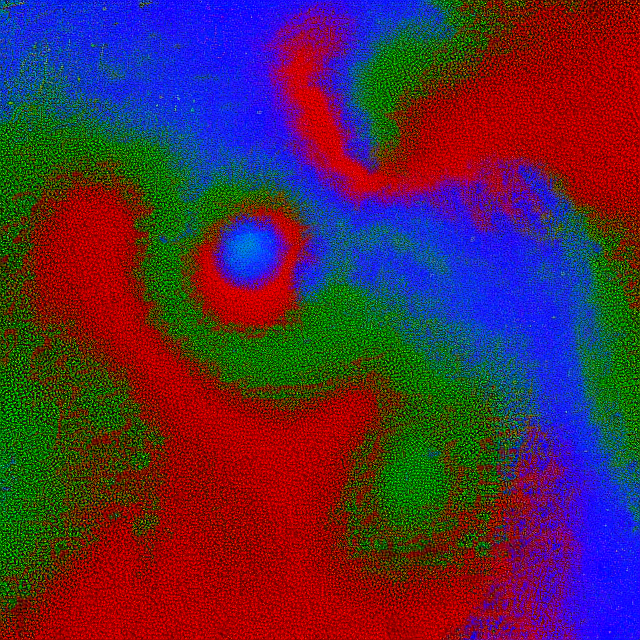} &
\includegraphics[width=0.30\textwidth]{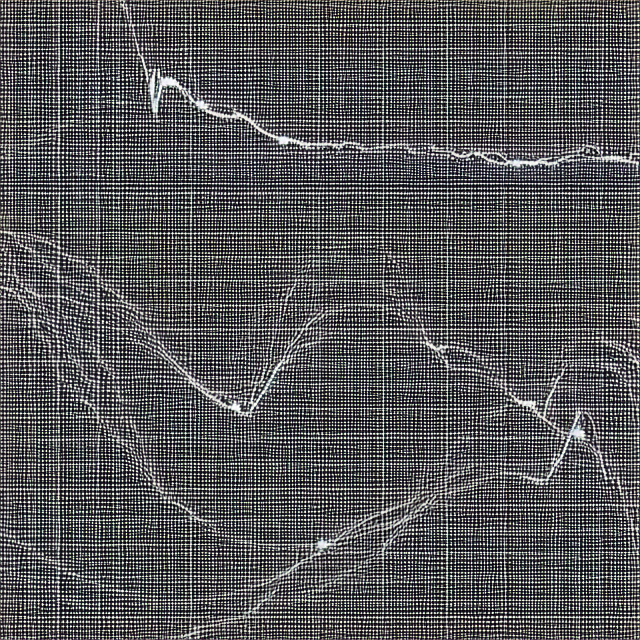} \\
\end{tabular}

\smallskip
{\small\itshape Fig.~01. Images produced by prompting Stable Diffusion 1.5 with ``Gaussian Noise.''}
\end{center}

The autoencoder is designed to reproduce the prior pipeline's ordering of the visible. What the autoencoder contributes is a frozen definition of visual plausibility. An image's symbolic function shifts: no longer a vessel of expression or communication, but information that helps a system pass quality checks. The image is verified not by what it represents but by what it resembles, enforcing a particular concept of what an image is meant to be and what kinds of images are desirable to the commercial market. Disembodied and without awareness, it accepts and returns a reductive symbolic violence (Bourdieu 1977). Esser and colleagues integrated this objective into the LDM via VQGAN (2020), carrying this hollowed-out perceptual position into the generative pipeline.

\subsection*{U-Net (Encoding)}

The U-Net was designed to examine the human body. Developed for biomedical microscopy (Ronneberger et al. 2015) and later used for tumor segmentation in MRI scans, it was deployed for the clinical encounter: a body made understandable to medicine, prone to misinterpretation. The question it aimed to answer (\emph{how do you detect an anomaly within a human body?}) was accountable to a specific person and specific cells taken from them in a particular hospital room. Ironically, the U-Net architecture was designed because of the small scale of this bodily specificity. Clinical datasets are small because annotation costs money, radiologists' time is limited, and ethical protocols restrict data collection. U-Net compresses the limited images available to it until only outlines are visible, then expands and restores spatial details through skip connections. U-Net found its place in identifying tumor boundaries without needing labeled images at every scale. Within this system, the body was referenceable, and it occupied the position of ground truth.

When engineers integrated the U-Net into the LDM, the specific body disappeared. The U-Net's task became denoising images from the commercial web. It progressively evaluates noisier images until it can find what noise had obscured, tracing an image back to its original form. But as engineers replaced the singular body as a referent with a vast sea of disconnected images, there is no ``original'' form for the U-Net to return to. What it recovers is the ideal representation inherited from the autoencoder's perceptual loss framework, the same frozen definition of visual plausibility built to pass automated quality checks. If the autoencoder replaced the human gaze, the U-Net is transformed by shifting what it gazes upon.

This is where training data comes to replace whatever the camera once pointed at. The result is a closed symbolic order, a system of differences with an ungrounded center. In \v{Z}i\v{z}ekian terms, even the gap has been structurally excluded from the model's relationship to the symbolic, which is constructed entirely of images that precisely align with textual descriptions. All else is filtered out before training begins. The clinical body contained a person, or in Lacanian terms, an unconscious structured by the gap between language and the desire the gap produces. Taking the body out of this circuit strips away that subject; it is the user who returns to fill it through the act of interpretation and requests. But the user is not \emph{seen}. The body removed from the circuit, the system responds exclusively to a symbolic order, shifting from corpse to corpus and transferring its accountability from an embodied subject to the filtered dataset. The structure of the LDM treats the symbolic as complete. It is unable to seek or demand anything that would lie beyond the image/text pair. The model operates on a closed symbolic order, akin to a structural psychosis. Only through our intrusion does the model find gaps, presented to us as errors and fissures in representation. The model has no sense of it either way.\footnote{Comparing the position of the image model to that of a language model, to produce text seems to produce authority, and to produce images seems to produce a fantasy.}

The visual corpora of the commercial web fill the structural position vacated by the body. This visual world is never directly encountered by the system because it is abstracted into latent geometry: inherited, pre-filtered, compressed, and determined, unlike the body it replaced. The skin the model reconstructs is not the anxious skin of a patient beneath a scanner --- it is a visual skin that earns upvotes on Reddit, the composition that passes aesthetic scoring, the figure that appears most plausible within the world of the commercial web. The site of accountability has moved from a human body to a body of e-commerce images.

\section*{Generation}

\begin{quote}
\itshape
We can now move away from the training of the model, which produces the latent space, to what the latent space then produces. This section grapples with the process of generating images, in which the previous paths make their marks, and in which new processes intervene.
\end{quote}

\subsection*{Tokenization and CLIP}

The prompt redefines the language we use, structuring ``what is written and how, what decisions can be made'' (Amoore 2025, 576). To prompt is to pre-translate desire into machine-legible terms. Within an LLM, the prompt is the seed that sprouts more language.\footnote{See also Salvaggio, ``The Fixed-Explosive: Language Models and the Blurry Subject'' (manuscript under review).} Within the diffusion model, the prompt becomes an anchor for a visual representation: the relationship between our prompt and the resulting image takes on a different flavor of abstraction, and a different position to its output. It converts a specific desire to \emph{see}, in a specific situation, into language for the model to break apart. The prompt does not enter the model as language. The text is first split into tokens --- small units that may be whole words, parts of words, or punctuation --- and each token is converted to a number. The system never reads the sentence; it reads the numbers. Steyerl notes that prompt-based generators train users to a vocabulary as much as they train models (2023). That training restructures the user's desire, orienting it toward what the token vocabulary can receive, and the cross-attention hierarchies can visualize: we prompt to meet the model's affordances. CLIP's tokenizer defines the vocabulary (what can enter the system at all), while CLIP's encoder determines what those tokens represent.

While earlier configurations of Latent Diffusion experimented with BERT-style text encoders (Rombach 2021; Devlin et al. 2018), the released versions adopted CLIP, the same mechanism used to sort the training corpus. CLIP's text encoder uses a 49,152-token vocabulary derived from its training corpus of approximately 400 million image-text pairs scraped from the web (Radford 2021). Words that appear more often in image-text contexts have dedicated tokens; everything else is broken into sub-word units, their specificity lost. Importantly, they enter the system already calibrated to visual co-occurrence by people linking to them; they are conditioned by what users of the commercial web have paired them with (see the Common Crawl section). CLIP's tokens are never parsed as pure language. The text encoder maps these tokens into a 768-dimensional vector space.\footnote{By ``dimensions,'' it means certain features of the text and image are each compressed into a list of 768 numbers which works as a coordinate in a shared space. If two coordinates are close together, the model considers the text and image more likely to be related.} There is no moment in the pipeline where the prompt exists as language before it becomes a vector trained on their value as visual commodity. Any social exchange value enters the system already converted through its financial value within the platform economy.

The interface presents this as natural language processing, as if the system heard the user. But text becomes vector before it ever becomes image --- a chain of abstractions that produces perpetual mistranslation. Mistranslation is not inherently negative. Of course, such gaps can frustrate users but can also delight through the discovery of unexpected images. Novelty arises from this failure to be precisely understood: you articulate yourself and the machine returns its flawed articulation. Without the gap there is no pleasure: the diffusion model would operate like a stock photo website. The process of generativity allows the user to close the gap, but the gap is instantaneously filled with a new proposition --- an invitation to adjust the prompt or adapt to what was provided. But in all cases, the system segments the user's expression of desire into units with fixed values --- activating a closed symbolic order that the user cannot inspect. The model cannot learn the user's desire; the user must adapt to the model's invisible boundaries.

Because text and image are entangled at the encoder level, the prompt cannot escape the visual associations that CLIP has already learned, and this comes back to the user. Here we find ideology inscribed into images most clearly. Bianchi, Kalluri, and colleagues found that ordinary prompts reliably produce stereotypes, with ``Africa'' associated with poverty and counter-modifiers like ``wealthy'' or ``fancy'' failing to disrupt the stereotype: ``an African man and his fancy house'' returns a hut, while ``a wealthy African man'' produces a man in a suit beside the same hut (Bianchi et al. 2023). Wealth could be pictured, but only as assimilation into the Western default. The findings suggest the model is structurally incapable of disentangling poverty from Blackness (Bianchi et al. 2023, 11--12).

One response to these findings is a call for more racially diverse training data or guardrails that correct demographic skew. But in diffusion models, the emphasis on data distracts from this structural flaw: no generated image can be \emph{accurate} because the architecture cannot verify it. A statistically correct racial distribution would not influence the model, which doesn't generate from that distribution but from the most represented composite. Visibility to the system is often not even desirable: deepfake pornography and information pollution, for example, are not problems of underrepresented demographics. Instead, the issue is the privileges afforded to users: a fantasy in which we are recognized and produce the images we desire. Failure misrepresents both the person prompting and whoever she imagines is depicted. The machine cannot meet us, because its symbolic order excludes any desire to do so. It returns to us as a stereotype structured by a compression of references.

\subsection*{The Transformer / Attention Mechanism}

Attention, the mechanism at the transformer's core, had answered a translation question: \emph{which words in a source sentence matter most for producing each word in the target sentence?} (Vaswani et al. 2017). Repurposed for image generation via cross-attention, it now answers a different one: \emph{which parts of a text prompt matter most for which regions of an image?} The mechanism is responsible for managing the compositional structure of the image in training and generation: that an eye is placed in relation to the nose and ear, rather than the image merely containing all of them in arbitrary arrangements.

The attention mechanism treats word-to-image translation as if it worked like language-to-language translation. It is based on the insight that natural language processing must parse context to generate reliable text, because ``bird'' has a different meaning beside ``brain'' than it does beside ``watching.'' For each part of the emerging image, it calculates which aspects of the user's prompt are most relevant to what should be depicted there. Where the tokenizer allows or blocks language into the image, the transformer prioritizes what is named. For example, the word ``fancy'' exists in the vocabulary but loses its significance against ``African man and house'' (Bianchi et al. 2023). The culture of prompt engineering reflects this hierarchy. Prompt engineers commonly add tokens like `masterpiece,' `trending on ArtStation,' and `cinematic lighting' to influence an image's style, speaking to a stylistic component of the image's final production. The user who learned to add them to every prompt discovered which tokens are most valued within the model's perceptual framework. Prompt engineering is, more accurately, \emph{prompt translation} --- an effort to interpret how the model converts language into visual representation.

The cross-attention hierarchy contributes to what Steyerl calls the mean image, which replaces ``likenesses with likelinesses,'' converging on the statistically dominant rather than the particular (Steyerl 2023). This does not visualize the internet's democratic average --- it visualizes the commercial internet's centralizing tendencies (Salvaggio 2022), compounded at every stage of the pipeline, and returned as visual plausibility. The user learns not only to want differently, but to see differently. The user is in negotiation with the attention mechanism, setting stray words as anchors for visual reference. This is part of the pleasure of the prompt engineer: the loading bar, the image's emergence, the dials that give her a sense of control over its response. The nature of the mechanism makes us feel like its failure to meet us is part of the joy. It also reinforces a specific orientation to vision, wherein style and meaning are articulatable in advance and categorizable through text, and recruits the user into a belief in the project of neural exchange. It claims that everything we see is commensurable, if only we can find the mechanism to translate it.

\subsection*{U-Net (Decoding)}

Within latent diffusion model's generation phase, the U-Net runs in reverse. It begins with random visual noise and removes that noise step by step, guided at each step by the prompt, until a coherent image appears. It operates in tension with the surrounding systems to identify the basic structural forms associated with the prompt (the shape of a house, or a body, or a squirrel). As each step is refined, it is measured against the remaining noise in the image to predict what should replace it. As it moves across these scales, the U-Net restores finer local details compared to the previous step: textures such as wood, skin, or fur, via skip connections.\footnote{Skip connections are shortcuts in the architecture in which information from the early compression stages are passed forward to the later expansion stages, so detail that would otherwise be lost can be restored via plausible approximations.}

U-Net is predicting the replacement of noise with particulars. These are details that describe what might exist at that scale for that basic structural shape, based on the prompt. The result is an image that appears to refer to something, assembled from a structure that refers only to itself. It is categorically correct --- we recognize it as a street scene or a face and can connect it to our request. What it lacks is particularity: what would make it a response to a specific desire rather than a generic instance of a type. The viewer senses this vacancy without being able to name it; colloquially, it takes on the aesthetic commonly described as ``slop.'' The image confirms that it belongs to a category but comes apart under sustained attention. These are the moments where the commodity form of the image becomes visible --- where a structure built from references to references finally shows through.

The image is a product of the noise that triggers the process. Noise in the system can be understood as clusters of pixels that do not represent a machine-recognizable object or detail. This absence of representational presence determines the structural form that the system fills in. At the start, it has no presence at all, converging quickly on rough outlines. The details follow the emergent structural form of that initial enclosure. The machine predicts where this visual noise remains and, at each step, how to replace it in service to the prompt. The machine thus transforms noise into signal through a regime of iterative filtering. It constrains noise until it becomes an image, constraining what does not fit its referential structure over a series of steps, adapting the scope of detail it can refine based on the number of steps it has been given.

Visual noise is the ultimate computational equivalent of representational variety in the system, occupying a space as a substitution for human creativity, with the U-Net operating in the role of discernment and taste, if it had to be equated to human image-making. But it exists only to be structured by the ``world within'' the diffusion model. It assesses the measure of noise against references drawn from a narrow, specific form of social communication, one absent from the experience of a body. But the very concept of structuring against noise creates a paradox. It suggests that any unrecognizable contingency within the image is the source of possibility, but at the same time, must be reinscribed entirely with references to its own closed world.

While models further on from latent diffusion have moved away from U-Net as the mechanism for this phase, the logic remains the same. Newer systems, such as Flux, or Diffusion Transformers, completely replaced the U-Net with the cross-attention mechanism to steer the denoising process, wherein the image-text pair exerts more control over each level of detail within the composition (Peebles and Xie 2022). Of course, the body remains displaced. Where U-Net was designed for a body but directed at a corpus of images, its replacement in the Diffusion Transformer models was designed for language but directed at the image: the body lost to language, a symbolic order with no body left inside it.

\subsection*{Classifier-Free Guidance}

Classifier-Free Guidance, or CFG, is the dial that controls how literally the model takes the prompt. When it is off, it simply produces the most likely image based on the noise seed it produces at that instant. Low values skew toward this statistical center, mediated by images that align with the prompt provided. At higher levels, tensions between the center and periphery are more pronounced and images become more unique to the prompted suggestion. At extreme levels, this distinction overemphasizes the prompt, saturating it while eliminating other context to the point of complete semantic collapse. Every generated image is steered by this dial, whether the user adjusts it or accepts the default.

CFG was developed as a cheaper alternative to classifier guidance, a distinct neural net outside of the diffusion model that compared and evaluated outputs during denoising toward the labeled category. That classifier was expensive to train and run. Classifier-free guidance trains a single model on the same data while randomly dropping the text prompt during training, so the one network learns to predict both with and without conditioning. At inference, the setting amplifies the difference between the two predictions, achieving comparable steering without added training cost (Ho and Salimans 2022). The result is a tunable dial, CFG scale, that controls how strongly the output is pulled toward the prompted prediction and away from the unprompted one. The question CFG appears to answer is \emph{how closely does this output match the prompt?} The question it really answers is \emph{how far is this image from the statistical center for the prompt?} Each increase in guidance strength enforces that distance more aggressively.

\begin{center}
\begin{tabular}{|>{\centering\arraybackslash}m{1.6cm}|>{\centering\arraybackslash}m{3.1cm}|m{6.6cm}|}
\hline
\multicolumn{3}{|p{12.0cm}|}{\textbf{``Commodity'' Generations in Stable Diffusion 1.5}\newline \itshape These images are a product of the same seed (the same structure of noise) in Stable Diffusion 1.5, prompted with the word ``commodity'' at various CFG settings.} \\
\hline
CFG 00 & \includegraphics[width=2.9cm]{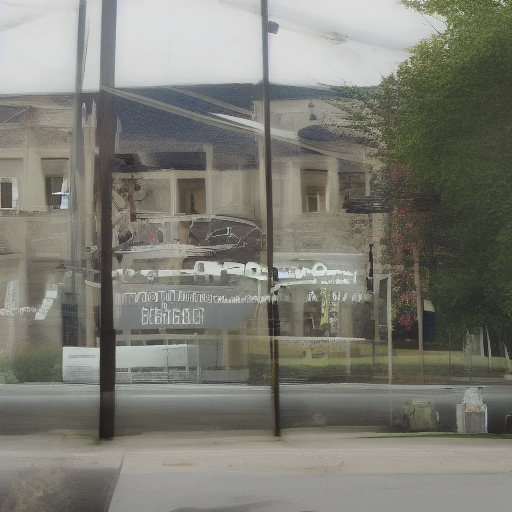} & \textbf{No Guidance.} Without guidance, the model draws from a statistical average of images from training. The prompt has no influence and any association with `commodity' is the user's. \\
\hline
CFG 03 & \includegraphics[width=2.9cm]{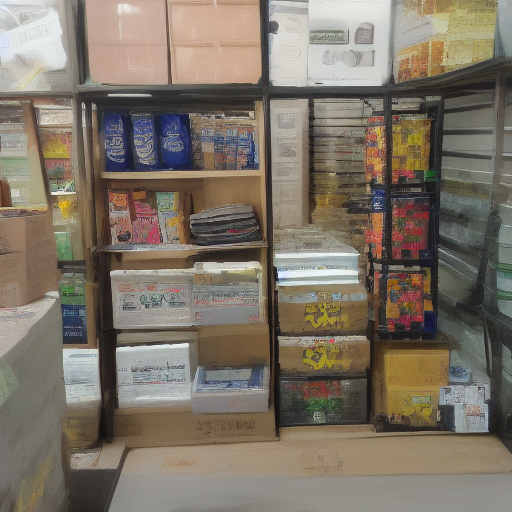} & \textbf{Low Guidance.} The prompt steers weakly toward the statistical center of the prompt. Shelves and packaging evoke the \emph{category} of commodities without arriving at any specific \emph{thing}. \\
\hline
CFG 12 & \includegraphics[width=2.9cm]{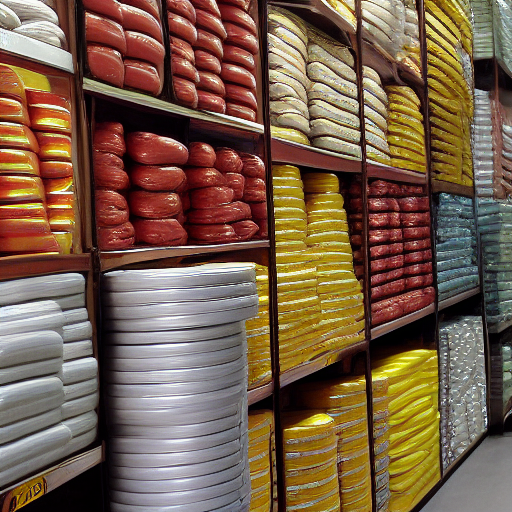} & \textbf{Average Guidance.} The model pulls toward the statistical center of its text-image pairs: shelves of packaged goods in saturated colors is the commodity form at its clearest. \\
\hline
CFG 25 & \includegraphics[width=2.9cm]{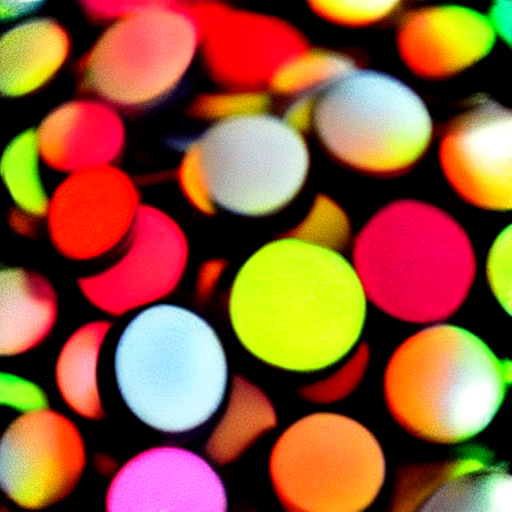} & \textbf{Extreme Guidance.} As each step amplifies the difference between prompted and unprompted predictions, the distance is pushed into incoherence. A reference to a visual structure without specificity. \\
\hline
\end{tabular}

\smallskip
{\small\itshape Fig.~02, an annotated series of images produced from the same seed at various CFG scales.}
\end{center}

The technique encodes a specific theory of control over the image. In neural-exchange terms, the dial sets the distance between two points: at the lowest end is the stacked center of a blurry composite, derived from the training data. At extreme higher settings this stack is stretched so thin as to become unrecognizable, distorting any anchors to the visual representations that rest inside that stacked center. In between these points on the dial is the range where commensurability can be most actively negotiated. Turn the dial up within that range, and the image moves toward the most typical, recognizable visual representation of the prompt. Turn it down; the image drifts toward what the model produces without instruction, the statistical center of the training distribution. What the user experiences as a control over fidelity is a control over commensurability: the degree to which the output conforms to the most exchangeable visual interpretation of the prompt.

The dial introduces control to the user on its terms, defined by how far the desired image sits from its internally assembled composite. If the user strays too far from commensurability, the image collapses entirely. At CFG 25, it pursues the prompt so aggressively that the output leaves the domain of recognizable images: the neon circles in Fig.~02 are what `commodity' looks like when the system's measure of fidelity is pushed beyond depicting anything at all. CFG 25 tells us something about every image the model produces: within the latent space, particularity has already been traded for commensurability --- each image is a position defined by its relations to all other positions, an abstract composite of images serving as a reference. CFG 25 visualizes what happens when an image is commensurable with nothing, yet still operating within the logic of commensurability.

\section*{Conclusion}

The design goal for the latent diffusion model is to produce a compelling mimicry, one that camouflages the distinction between the model and its operator into a homophilic resemblance. To create this illusion, the diffusion model is presented as an object without seams. The reading offered here treats it as an ensemble of mechanisms, each performing its own role. The camouflage of intelligence and creativity ascribed to these models masks the automated, large-scale compression of the social sphere into vectoral representations within enclosed, self-referential neural networks. We examined training as the site of this translation: within the neural economy, data is valuable not for its uniqueness but for reinforcing plausible ties between word and image. With human social exchange as its raw material, data's value shifts from scarcity to substitutability: we rarely discuss \emph{datum}, but always \emph{data}, the collective noun, as if any individual data point has no individual value.

What is lost in this conversion is the groundedness that anchors the particular within the world. Social exchanges are accountable to something beyond themselves: an archive of correspondences and photographic traces that tie to embodied experiences and leave a remainder. The model loses that anchor when the system's gaze is shifted away from the body,\footnote{The displaced body is a consistent motivator of Silicon Valley's TESCREAL ideology; see Gebru and Torres 2024.} from the corporeal to the corpora. The U-Net closes in on the ranked image, operating at an infrastructural distance even from the image that replaces the body. It is limited to a closed, compressed, and amplified symbolic order that mimics the forms of social exchange in isolation from the social sphere that allows these images to function.

The analysis of the model architecture proposed in this paper enables more precise questions and vocabulary about the mechanisms within the ``seamless'' object of the generative AI model. While focused on the latent diffusion model, the approach emphasizes an analysis of automation that asks \emph{what decisions are being automated} and \emph{what logic is inscribed within those mechanisms to guide them}. The aim is to broaden, not abandon, the emphasis of biases and ideologies in datasets to include the decisions about how these systems transform that data. This calls for us to escape the frame of the ``AI'' system as a singular ``model,'' to excavate and examine the origins of its components and tasks; and to acknowledge the ideological power involved in building structures of automated decisions. To do so opens medium-specific lenses for interpreting synthetic images and what they do in the world.

The poet Paul Val\'ery suggests that the poem is a tool for conveying feeling: not intended to transfer the writer's inner experience to the reader, but to serve as a platform through which the reader may \emph{reconstruct the source} of the feeling (1985, 60). Pinning language to the experience of the world is an exercise in perpetual misnaming. Yet the image returns in response to the language used to call it forth; the user then contorts themselves to whatever arrives.\footnote{Salvaggio, ``The Fixed-Explosive: Language Models and the Blurry Subject'' (manuscript under review).} We ought to ask who these images are from and for and what we are meant to reconstruct. Every stage of image generation inherits an absence of subjectivity and an inability to imagine --- and therefore ever to \emph{meet} --- the user. The generated image displaces the social labor of making and circulating an image; it produces the commodity for its own sake. The image, in pure \emph{commodity} form, becomes the object desired by the prompter, allowing the user to enjoy the role of creator \emph{and} to enjoy their alienation from its production. As a fetish object, the generated image bears only sparse traces of its author's expression --- a prompt, a handful of settings --- yet the author can still redistribute it alongside a desired interpretation. In contrast to the generated image, the poetic expression of an inner state must seem terrifyingly intimate. The generated image, as a fetishized commodity of the platform economy, allows us to assert and disavow the pleasure of social exchange without extending our vulnerability to the social world. It jumps ahead to create an image optimized for the conditions of online digital circulation, regardless of what it communicates: its audience is, foremost, the algorithmic systems of the social media platforms from which it is derived.

In contrast, the critical resistance to AI risks a fixation on the return to the logic of physical commodities. It reaches for commodity defenses --- copyright law, data sovereignty, consumer boycotts --- engaging the neural economy not by centering social exchange but by insisting the image is a marketable good rather than a vessel for communication in a digital culture. These defenses are not wrong. They resort to positions where economic policy and the law may intervene, but do not address the alienation arising from the constant reduction of the social world to total commensurability. The market \emph{around} the model looks to commercialize this logic by promising automation of digital commodities. The market \emph{within} the model dehumanizes through the abstract compression of the social sphere. These are distinct but intertwined positions. Without acknowledging this, critique risks reliance upon the logic of commercial markets. Platforms, meanwhile, use images to mediate social exchange in ways that go beyond economic value, accelerating cognitive and platform capitalism.

The system scraped our participation in social exchange; what it returns is the artifact of participation without the relationships that produced it. This is what makes generative AI simultaneously intimate and repellent. The neural economy absorbs us as subjects with desires, intentions, and an urge to resist. We resist it as subjects for whom approximation feels like desire and commodification feels like resistance. The model is made by flattening the bumpy subjectivities within human communication into an algorithmic pin-board for commodification of its by-products. So, increasingly, are we. If we do not place participation in the social sphere at the center of the critique, we risk reaffirming the very fetishism of the commodity the model produces.

For now, these systems are positioned to process information from a diminished world, narrowed by a particular ideology of vision: images mediated by platform and attention economies, measured by their suitability for literal description, become subjected to a medical gaze that was once aimed at finding symptoms within a literal body. The user of these systems conjures these symptoms as images but is offered only perpetual misdiagnosis: they believe they see the social world of image circulation and communication. But in the place of the body is a market whose transactions are obscured under the guise of images and creativity.

\section*{Acknowledgments}

This research was supported through PhD funding from the Gates Cambridge Trust. Special thanks to Leonardo Impett, Caroline Bassett, Ryan Heuser, and Aline Guillermet for notes and draft feedback.

\end{document}